# Mechanistic Insights into the Hydrazine-induced Chemical Reduction Pathway of Graphene Oxide

Shu Chen, [§a] Jianqiang Guo [*§b]

[a] Department of Physics and Astronomy, Northwestern University, Evanston, IL 60208, USA

[b] AECC Beijing Institute of Aeronautical Materials, Beijing 100095, China

**Abstract**

Hydrazine stands out as the most generally used chemical-reducing agent for reducing graphene oxide. Despite numerous experimental and theoretical investigations into the reduction reaction, the reduction mechanism remains unclear. In this study, we propose that, in aqueous hydrazine solutions, both hydrazine and hydroxide ions could initiate the reduction of graphene oxide. We introduce a chemical reaction pathway involving C-H cleavage and a dehydroxylation process for the reduction of graphene oxide. By utilizing density functional theory calculations, the reduction reactions mediated by hydrazine and hydroxide ions are separately investigated. The reaction routes on the basal plane and edge regions of graphene oxide are discussed independently. The density functional theory calculations demonstrate that the proposed mechanism is both thermodynamically and dynamically feasible. This work might contribute to an atomic-level comprehension of a longstanding challenge in the field of graphene oxide.

# 1. Introduction

Graphene has received considerable attention since its discovery in 2004 due to its fascinating properties.[1] As a derivative of graphene, graphene oxide (GO) integrates a variety of oxygen-containing functional groups, as well as $sp^2$ conjugated regions and holes.[2, 3] The chemical reduction of GO is considered to be the most reliable and feasible method for mass-scale production of graphene materials because of its cost-effectiveness and ease of operation.[4] To date, about 60 chemical-reducing agents have been applied for the reduction of GO.[4, 5] Despite the widespread use of chemical reduction strategy in research and industry, the mechanism behind the reduction of GO is not yet fully understood.

Hydrazine has been the most generally used chemical reductant for reducing GO among the various methods available.[6, 7] The hydrazine-reduced graphene oxide exhibits significantly improved electrical and structural properties, which are close to those of pristine graphene from the perspective of the chemical reduction method.[8] Numerous experimental studies and theoretical calculations have been conducted to investigate the hydrazine-mediated reduction reaction. In 2012, Park et al. performed a detailed characterization of $^{13}$C-labelled GO that had been treated with $^{15}$N-labelled hydrazine.[9] The solid-state NMR spectra revealed almost complete removal of epoxy and hydroxyl groups, the conversion from carbonyl to pyrazole, and trivial changes in carboxyl content. Pumera and co-workers conducted a reaction model approach to investigate the hydrazine-mediated reduction of GO in 2015.[10] Their research shows that the hydroxyl and carboxyl groups were not readily removed, while carbonyl groups formed corresponding hydrazone complexes.

It is worth pointing out that the reduction of GO is typically associated with the elimination of oxygen-containing groups and formation of new C=C bonds. Hence, the conversion from C=O (carbonyl) to C=N (pyrazole and/or hydrazone) is more of a "substitution" reaction rather than a "reduction" reaction. Additionally, GO only contains trace amounts of carbonyl and carboxyl groups at the edges and defects. Therefore, research on the chemical reduction of GO mainly focuses on hydroxyl and epoxy groups.

Standkovich et al. proposed a reduction pathway for the epoxy moiety, as shown in Scheme 1.[6] According to this reaction pathway, hydrazine reacts with epoxy groups to form hydrazine alcohols and aminoaziridine moieties, which could then undergo thermal elimination of diimide to form C=C bonds. However, the exact reaction path for the formation of the C=C double bond from the aziridine ring was not described in detail. Nagase and co-workers subsequently used DFT calculations to verify this reaction mechanism,[11] but they did not find the mechanism described in Scheme 1 and instead proposed three new reaction pathways for hydrazine-mediated reduction of epoxide groups located on GO. Furthermore, no clear reaction paths have been identified for the reduction of hydroxyl groups.

Kim et al. examined the epoxide reduction mechanism with hydrazine using density functional theory (DFT).[12] Their results showed that the epoxide ring-opening reaction tended to occur spontaneously at room temperature, with a low barrier of 10.2 kcal/mol.[12] Given that the epoxy groups will also be converted into hydroxyl groups during the reduction process, the key issue for the hydrazine-mediated chemical reduction of GO is essentially the reduction of hydroxyl groups. However, to our knowledge, the reduction mechanism of hydroxyl groups is currently not well understood.

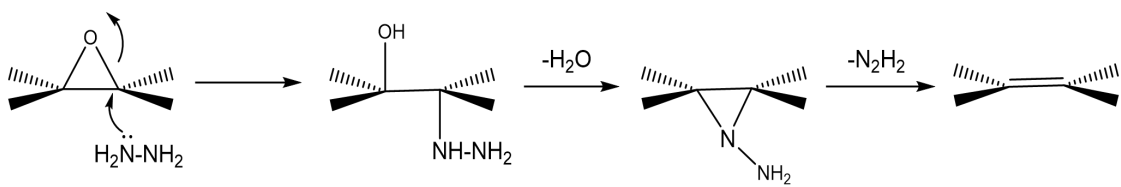

**Scheme 1 Reaction pathway proposed in reference XX for the reduction of epoxy with hydrazine**

In view of this, we propose an atomic-level reaction mechanism for the chemical reduction of hydroxyl groups. Theoretical calculations have been carried out for the hydrazine-mediated reduction of hydroxyl groups using density functional theory, and the results indicate that our proposed mechanism is favorable in terms of kinetics and thermodynamics.

## 2. Material and Methods

2.1 GO model

Due to the amorphous nature of graphene oxide, creating a standard model has always been a challenge. Graphene oxide contains many $sp^2$ and $sp^3$ hybrid regions, along with abundant oxygen-containing groups. The C/O ratio of GO typically falls within the range of 1-2.5, depending on the preparation method and the degree of oxidation. To simplify the problem, we established a model based on the coronene [$C_{24}H_{12}$] molecule. Given that the edge and basal plane regions of GO often exhibit distinct reactivity, we established models with hydroxyl groups located in the basal plane and at the edge, respectively.

2.2 Computational details

Quantum chemical calculations were carried out by using density functional theory (DFT) implemented in GAUSSIAN 16 program.[13] The structure optimization and single point energy were performed at M06 2X-GD3/Def2svp level. All structures were fully optimized with the solvent effect corrected by using the integral equation formalism polarizable continuum model (IEFPCM)[14] and the SMD radii[15] for water. We calculated thermal corrections for all optimized structures within the harmonic potential approximation at 298.15 K and 1 atm pressure, ensuring that all intermediates had no imaginary frequencies and that each transition state had only one imaginary frequency. In addition, we performed intrinsic reaction coordinate calculations at the same level of theory for each reaction path to confirm that the transition states were properly connecting the reactants and products.

2.3 Materials and method

GO was prepared using a modified Hummers' method. All the reagents were purchased from Aladdin-reagent Inc. and used as received. The chemical structures of the GO samples were characterized by Fourier-transform infrared spectroscopy (FTIR, Nicolet 6700).

3. Results and discussion

It is well accepted that the chemical reduction of GO involves the elimination of oxygen-containing functionalities and the simultaneous formation of new C=C double bonds. Therefore, in the pursuit of elucidating the pathway for reducing epoxy/hydroxyl groups, it's reasonable to hypothesize their elimination, in conjunction with adjacent functional groups, would facilitate the formation of C=C double bonds. In the case of hydrazine-mediated reduction of GO, previous research so far has primarily focused on the formation of C=C double bonds based on C-N (C-$NHNH_2$) cleavage,[4] as shown in Scheme 2a. However, in the network structure of GO, each C-OH may be connected to up to three adjacent carbon atoms (Scheme 2b). This means there exist three potential positions for forming new C=C bonds. Consequently, a new C=C bond may also be formed through the breaking of other adjacent $sp^3$ hybridized bonds, such as C-O and C-H.

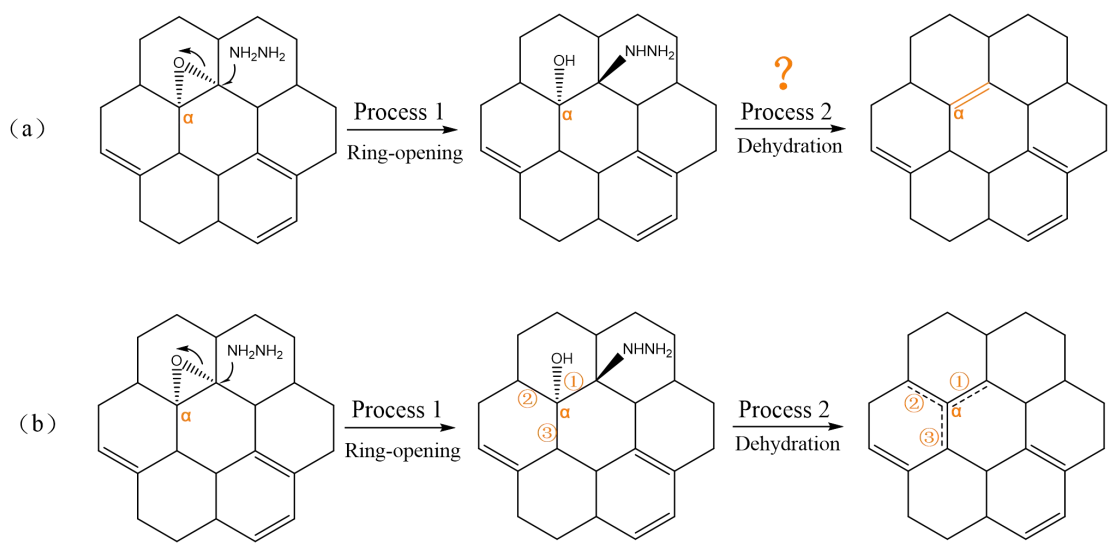

**Scheme 2** Reaction pathway for the reduction of epoxy with hydrazine

In contrast to oxygen-containing groups, there has been limited focus on C-H species within GO sheets. The presence of C-H bonds can be highlighted using FTIR spectra, as illustrated in Figure 1. The FTIR spectra of GO powder reveal two peaks at 2925 and 2855 cm$^{-1}$, corresponding to C-H stretching bands.

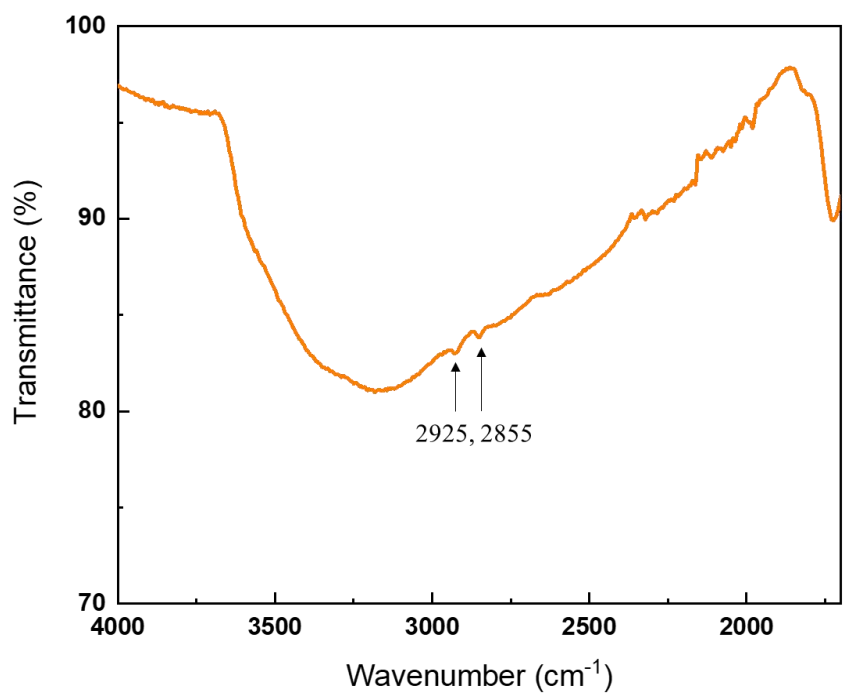

**Figure.1 FTIR spectra of GO powder**

Notably, as early as 2012, Kim and co-workers reported the presence of C-H moieties in GO, demonstrating their ability to facilitate the reduction of GO even under ambient conditions.[16] They discovered that the presence of C-H species in GO contributes to the reduction of epoxides and hydroxyl groups. According to their theoretical calculations, C-H is easily broken and initiates an H-mediated reduction reaction, eventually leading to the gradual reduction of GO at room temperature.

Given the alkalinity of hydrazine in its aqueous form, there is a strong likelihood that it will promote the activation of C-H bonds, thereby facilitating the reduction of hydroxyl groups. With this in mind, we propose a reaction pathway for hydrazine-mediated hydroxyl reduction, as shown in Scheme 3.

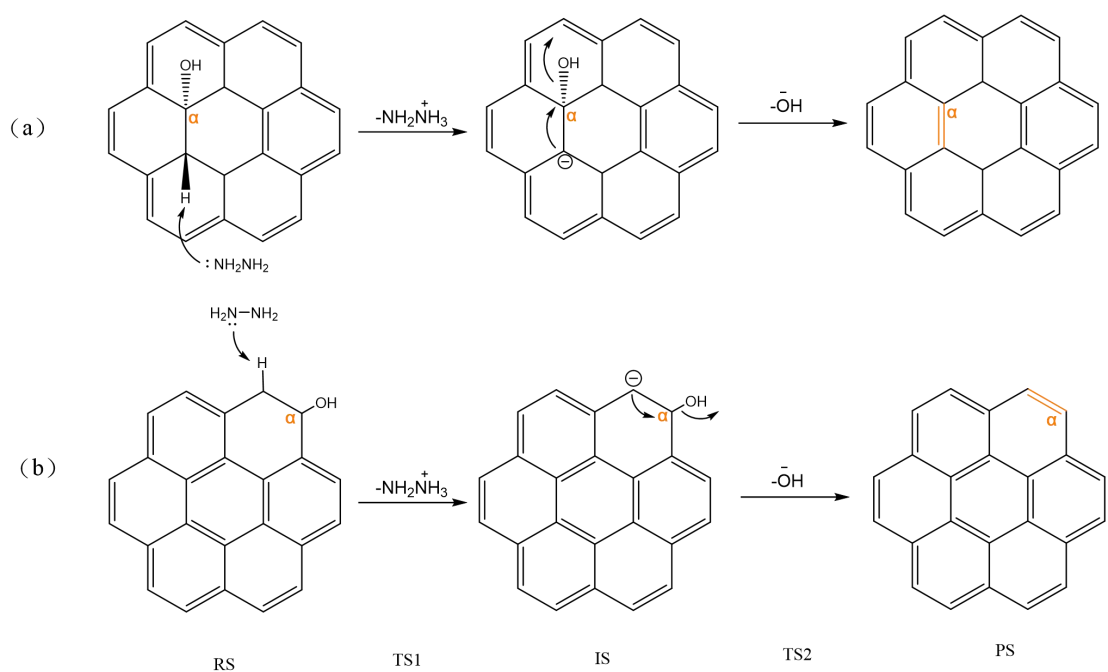

**Scheme 3 A proposed reaction pathway for hydrazine-mediated reduction of hydroxyl groups, (a) basal plane, (b) edge plane**

First, hydrazine attracts the H atom and drives the C–H bond cleavage, generating a carbanion intermediate. Subsequently, the electron pair on the carbanion displaces the hydroxyl group, forming a C=C bond. Considering the difference in reactivity between the edge and the basal plane regions of GO, we conducted theoretical calculations of the hydroxyl reduction reactions at the two positions, respectively.

Figure 2 shows the reaction energy profiles for hydrazine-mediated reduction of GO at room temperature. The activation energies required for cleaving the C-H bond on the basal plane and at the edge (TS1) are 4.87 and 27.89 kcal/mol, correspondingly. This result suggests that the C-H bonds within the basal plane are more prone to cleavage, while those at the edges exhibit lower reactivity. The activation energies required for breaking C-OH bonds on the basal plane and edges (TS2) are 18.18 (15.96 - (-2.22)) and 30.20 (58.63 - 28.43) kcal/mol, respectively, which indicates that the edge hydroxyl group has lower reactivity compared to its counterpart on the basal plane. Furthermore, the energy needed to break the C-OH bonds in both regions exceeds the energy required for cleaving the C-H bonds. This implies that in both scenarios, the rate-determining step is the cleavage of the C-OH bond.

Additionally, the reduction of hydroxyl groups on the basal plane yields a Gibbs free energy change (ΔG) of -58.25 kcal/mol, while at the edge, it's -11.52 kcal/mol. This implies that the reduction on the basal plane is more thermodynamically favorable. It can be concluded that hydroxyl groups on the basal plane exhibit a tendency for reduction by hydrazine at room temperature, driven by their lower activation energy and more negative ΔG. In contrast, hydroxyl groups at the edge are less conducive to reduction, primarily owing to their higher activation energy.

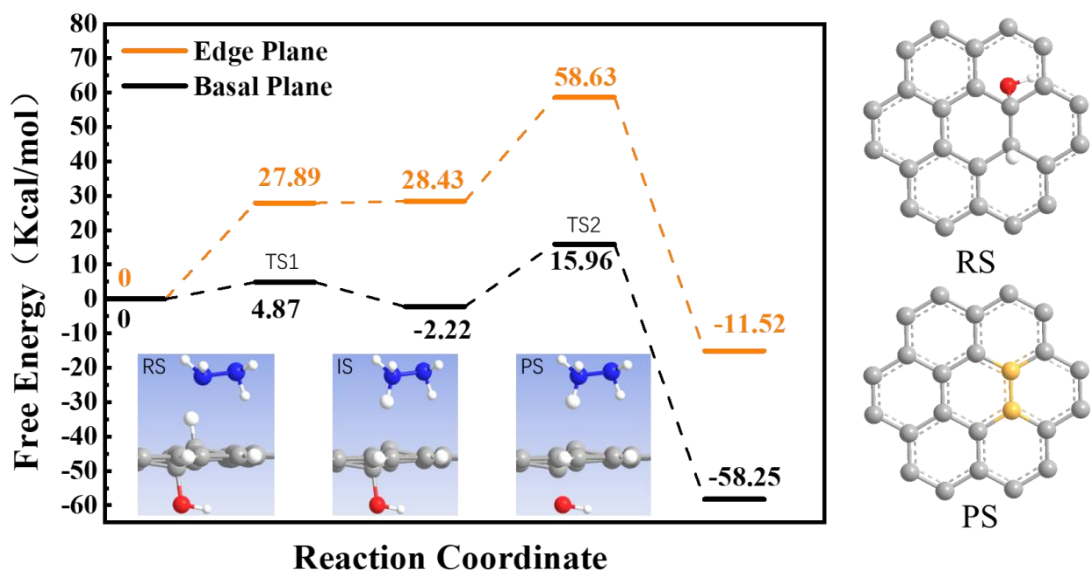

Figure.2 Reaction profiles for hydrazine-mediated reduction of hydroxyl group. All energies (in kcal/mol) are relative to the reactants. Color scheme: C, grey; O, red; N, blue; H, white. RS: reactants; TS: transition state; IS: intermediate state; PS: products.

It is worth emphasizing that the above-mentioned reaction pathway is based on the premise that hydroxyl groups are adjacent to C-H bonds. Therefore, it is necessary to examine the scenario in which the C-H bond is not situated at the ortho hydroxyl position. To further explore the reaction dynamics in such cases, we established a comparative GO model featuring a δ C-H bond, as shown in Scheme 4a.

According to the above-mentioned theoretical calculations, the energy required for C-H cleavage within the GO basal plane is a mere 4.87 kcal/mol, suggesting easy cleavage resulting in the generation of a carbanion. Then, the carbanion can be delocalized to form a stable resonance structure, as shown in Scheme 4a(II). It is found that, via this reaction pathway, structures a and b can yield an identical resonance structure (Scheme 4a(II) and 4b(II)), indicating their equivalence in state. Ultimately, despite the variation in C-H positions, the carbanions in Scheme 4a and 4b produced identical reduction products owing to the resonance structure. From another perspective, it appears that hydrogen atoms in C-H bonds are diffusing along

the conjugated network in GO. This result is consistent with the report by Kim et al.[16]

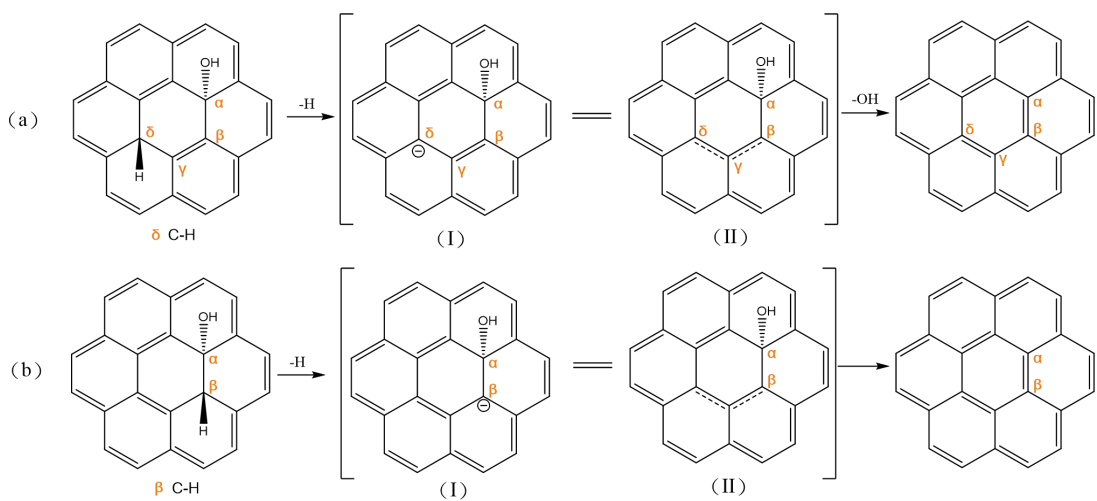

**Scheme 4** A proposed reaction pathway for the reduction of hydroxyl groups with a (a) δ C-H bond, and (b) β C-H bond.

The alkalinity of hydrazine promotes the cleavage of the C-H bond, initiating the reduction of hydroxyl groups. In the aqueous solution of hydrazine, hydroxide ions produced by the hydrolysis of hydrazine may also facilitate the breaking of the C-H bond. Thus, it is necessary to investigate the hydroxide ion-mediated reduction of hydroxyl groups and compare it with the reduction mediated by hydrazine. This exploration may also provide a detailed atomic-level mechanistic understanding of the reduction of GO by various alkaline reagents.

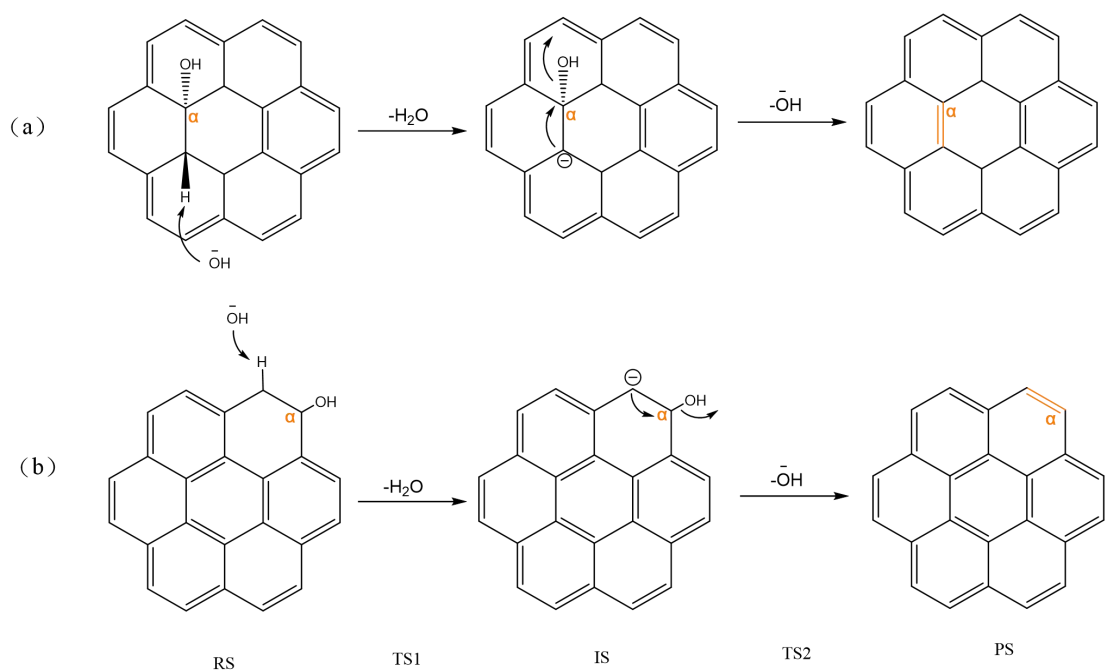

**Scheme 5** A proposed reaction pathway for hydroxide ion-mediated reduction of hydroxyl groups, (a) basal plane, (b) edge plane

As shown in Scheme 5, a hydroxide ion attracts the H atom, promoting the cleavage of the C–H bond and yielding a carbanion intermediate. Following this, the electron pair on the carbanion expels the hydroxyl group, resulting in the formation of a C=C bond. Theoretical calculations were conducted to explore hydroxyl reduction reactions on the basal plane and at the edge separately

Figure 3 shows the reaction energy profiles for hydroxide ion-mediated reduction of GO at room temperature. The cleavage of C-H bonds on the basal plane and at the edge (TS1) occurs easily, featuring activation energies of -8.52 and -11.14 kcal/mol, respectively. This implies both C-H bonds will promptly cleave and transform into carbanion intermediates. The activation energies required for breaking C-OH bonds on the basal plane and edges (TS2) are 17.05 (-53.78 – (-70.83)) and 23.59 (-11.45 - (-35.04)) kcal/mol, respectively. The reduction of hydroxyl groups on the basal plane appears to occur more readily. Similarly, in both scenarios, the cleavage of C-OH constitutes the rate-determining step for the hydroxide ion-mediated reduction reaction.

Additionally, the reduction of hydroxyl groups on the basal plane yields a ΔG of -54.35 kcal/mol, while at the edge, it's -11.52 kcal/mol. This implies that the reduction on the basal plane is more thermodynamically favorable. Despite both reaction pathways displaying negative ΔG, their free energies are higher than those of the carbanion intermediates. This suggests that the dehydroxylation process is thermodynamically challenging at room temperature.

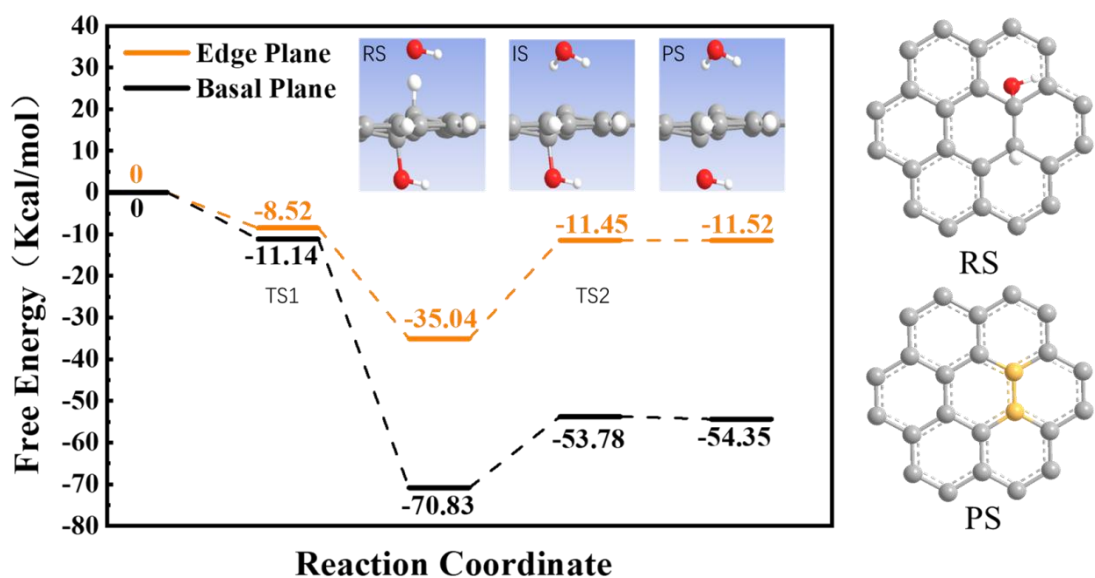

**Figure.3 Reaction profiles for hydroxide ion-mediated reduction of hydroxyl group. All energies (in kcal/mol) are relative to the reactants. Color scheme: C, grey; O, red; H, white. RS: reactants; TS: transition state; IS: intermediate state; PS: products.**

4. Conclusion

In this study, we suggest that the key issue for hydrazine-mediated reduction of GO is the reduction of hydroxyl groups. In aqueous hydrazine solutions, both hydrazine and hydroxide ions could promptly initiate the reduction of GO. We propose a chemical reaction pathway involving C-H cleavage and dehydroxylation for the reduction of GO. DFT calculation results show that the proposed mechanism is both dynamically and thermodynamically favorable. The cleavage of C-OH stands as the rate-determining step for hydrazine and hydroxide ion-mediated reduction reaction. In comparison to hydroxyl groups located at the edge of GO, those on the basal plane exhibit higher reactivity and are more easily reduced. This study contributes to the development of an atomic-level understanding of hydrazine-mediated reduction in GO.